\documentclass[aps,prl,twocolumn,amsmath,amssymb,showpacs,superscriptaddress,notitlepage,longbibliography,10pt]{revtex4-1}

\usepackage{graphicx}
\usepackage{dcolumn}
\usepackage{bm}
\usepackage{hyperref}
\hypersetup{colorlinks, linkcolor={blue}, citecolor={blue}, urlcolor={blue}}
\usepackage{xcolor}
\usepackage{mathrsfs}
\usepackage[T1]{fontenc}
\usepackage[utf8]{inputenc}
\usepackage{textcomp}

\begin{document}
	

\title{Anisotropic Surface Spin Waves as Signature of A-type Altermagnets}

\author{Zhoujian Sun}
\affiliation{State Key Laboratory of Quantum Functional Materials, Department of Physics, and Guangdong Basic Research Center of Excellence for Quantum Science, Southern University of Science and Technology (SUSTech), Shenzhen 518055, China}

\author{Yiyuan Chen}
\affiliation{Quantum Science Center of Guangdong-Hong Kong-Macao Greater Bay Area (Guangdong), Shenzhen 518045, China}

\author{Tao Yu}
\affiliation{School of Physics, Huazhong University of Science and Technology, Wuhan 430074, China}

\author{Hai-Zhou Lu}
\email{Corresponding author: luhz@sustech.edu.cn}
\affiliation{State Key Laboratory of Quantum Functional Materials, Department of Physics, and Guangdong Basic Research Center of Excellence for Quantum Science, Southern University of Science and Technology (SUSTech), Shenzhen 518055, China}
\affiliation{Quantum Science Center of Guangdong-Hong Kong-Macao Greater Bay Area (Guangdong), Shenzhen 518045, China}

\author{X. C. Xie}
\affiliation{International Center for Quantum Materials, School of Physics, Peking University, Beijing 100871, China}
\affiliation{Institute for Nanoelectronic Devices and Quantum Computing, Fudan University, Shanghai 200433, China}
\affiliation{Hefei National Laboratory, Hefei 230088, China}

\date{\today}

\begin{abstract} 
 Altermagnets have attracted intense interest because they have the advantages of both ferromagnets and antiferromagnets. However, their experimental identification remains challenging, in particular for the A-type altermagnets that account for a large group of material candidates. Here, we discover a kind of anisotropic surface spin waves in A-type altermagnets, which is absent in ferromagnets and conventional antiferromagnets.
 The anisotropic surface spin waves arise directly from the nature of altermagnets, i.e., the spin-opposite sublattices cannot be related by translation or inversion, which breaks the combined spatial-inversion and time-reversal symmetry, leading to the anisotropic surface spin waves with two properties, the chirality-dependent top-bottom positions and chiral split constant frequency contours. We further show that these two properties can be measured experimentally from the stray field and by resonance absorption spectrum, respectively. Our results provide a signature for detecting altermagnets and will inspire spin-based logic and information-storage devices.
\end{abstract} 
\maketitle

\textit{\color{blue}Introduction.---} Altermagnets are distinct collinear antiferromagnets with spin-opposite sublattices 
not related by inversion or translation~\cite{ PhysRevX.12.040501,PhysRevX.12.031042,fedchenko2024observation, chen2025unconventional,PhysRevLett.131.256703}. They combines the advantages of both ferromagnets and conventional antiferromagnets, making them promising in spintronics applications. On one side, they have rapid magnetic dynamics like conventional antiferromagnets, so writing them is much faster than ferromagnets~\cite{wadley2016electrical,vaidya2020subterahertz,park2011spin,PhysRevLett.109.137201,marti2014room,chen2025electrical,PhysRevLett.134.106801,PhysRevLett.134.106802}. More importantly, unlike conventional antiferromagnets, they could be read out via the anomalous Hall effect like ferromagnets, because their unique spin-opposite sublattices break time reversal symmetry~\cite{PhysRevX.12.021016, naka2019spin,PhysRevB.102.075112}. Therefore, searching for materials of altermagnetism has attracted great interest~\cite{bai2024altermagnetism,song2025altermagnets,jungwirth2025altermagnetism}, leading to the predictions of fascinating phenomena~\cite{PhysRevLett.133.056401,PhysRevLett.132.263402,jiang2024discovery,PhysRevB.108.184505,PhysRevLett.132.056701,PhysRevB.109.L201404,PhysRevLett.131.076003,PhysRevB.110.054427,PhysRevB.108.L140408,PhysRevLett.133.196701,PhysRevB.111.165406,PhysRevB.111.125119,PhysRevB.111.045409,PhysRevB.111.L121401}, such as giant tunneling magnetoresistance ~\cite{PhysRevX.12.011028,shao2021spin}, electrical spin splitter~\cite{ma2021multifunctional,PhysRevLett.126.127701,bose2022tilted}, finite-momentum Cooper pairing~\cite{zhang2024finite,PhysRevResearch.5.043171,PhysRevB.110.L060508,PhysRevB.111.054501,b7rh-v7nq}, and Coulomb drag~\cite{PhysRevLett.134.136301}, as well as increasing experimental explorations, including the angle-resolved photoemission spectroscopy \cite{krempasky2024altermagnetic,PhysRevB.109.115102,PhysRevLett.132.036702,zhu2024observation,zeng2024observation,yang2025three,PhysRevLett.133.206401,li2025topological,zhang2025crystal,jiang2025metallic}, $\text{X}$-ray magnetic circular dichroism \cite{amin2024nanoscale,hariki2024x,galindez2025revealing,cm39-hxqk,yamamoto2025altermagnetic}, transport \cite{feng2022anomalous,PhysRevLett.130.036702, PhysRevLett.133.106701, PhysRevB.110.094425,tschirner2023saturation,PhysRevB.110.155201,reichlova2024observation, gonzalez2024anisotropic,badura2025observation,rv1n-vr4p,PhysRevLett.133.056701,PhysRevLett.130.216701,g5xq-z15c}, and magnon spectroscopy \cite{liu2024chiral, 7yhz-jptc,singh2025chiral,biniskos2025systematic,512v-n5f9}. In particular, 
A-type altermagnets have intralayer ferromagnetism and interlayer antiferromagism~\cite{PhysRevLett.133.166701,dale2024non,sakhya2025electronic,regmi2025altermagnetism,zhou2025manipulation,reimers2024direct,PhysRevB.110.174410}, constituting a large group of material candidates. However, the A-type altermagnets with out-of-plane N\'{e}el order forbid the anomalous Hall signal~\cite{wang2023emergent}, making them hard to be detected.



\begin{figure}[htp!]
	\includegraphics[width=0.48\textwidth]{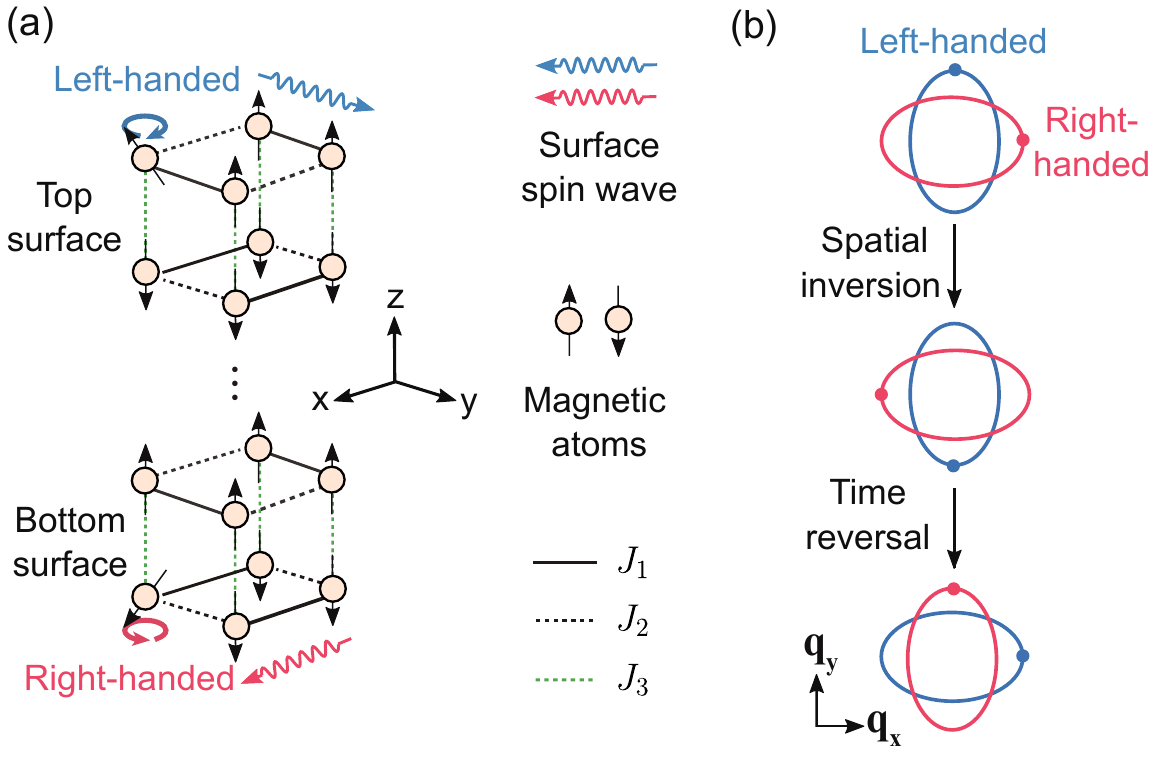}
	\caption{Schematic illustration of the anisotropy of the surface spin waves in an A-type altermagnet as a result of the intrinsic $\mathcal{PT}$-symmetry breaking ($\mathcal{P}$ for spatial inversion, $\mathcal{T}$ for time reversal). (a) When $J_1 \neq J_2$, the interlayer spin-up and spin-down sublattices are no longer connected by translation or inversion,  which satisfies the definition of altermagnet and breaks the $\mathcal{PT}$ symmetry, as illustrated by the Constant-frequency contours of the surface spin wave dispersion on the $q_x-q_y$ plane in (b). As a result of the broken $\mathcal{PT}$ symmetry, a left-handed (right-handed) surface spin wave can be found only propagating along the $y$ ($x$) direction and localized on the top (bottom) surface at a certain frequency. This anisotropy is absent in other systems, thus could serve as a compelling signature for detecting altermagnetism.} 
	\label{Fig1}	
\end{figure}

\begin{figure*}[htbp]
	\includegraphics[width=\textwidth]{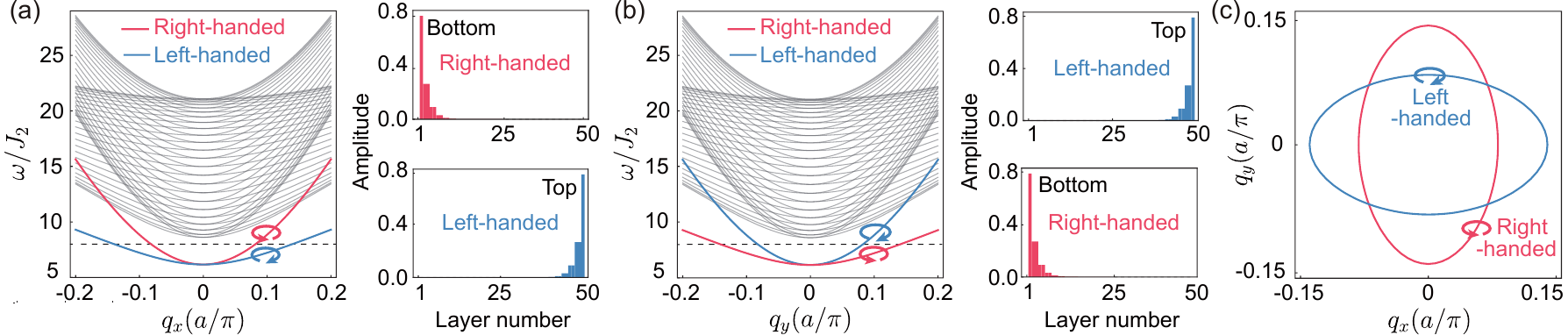}
	\caption{Dispersion of the spin waves in the A-type altermagnet as functions of (a) the $x$-direction wave vector $q_x$ and (b) $y$-direction wave vector $q_y$. $\omega$ is the frequency of the spin waves in units of the ferromagnetic coupling $J_2$. The gray curves are the spin waves from the bulk. The red and blue curves are two surface spin waves with the right-handed and left-handed chirality, respectively. 
    The right panels of (a) and (b) show the amplitudes of the surface spin waves at the energy marked by the dashed line as functions of the layer number (1 for bottom surface and 50 for top surface). (c) Equal-frequency contours in the $(q_x,q_y)$ plane at $\omega/J_2=8.0$ (dashed lines in (a) and (b)), showing the anisotropy of the right-handed and left-handed surface waves. The parameters are $q_{x,y}=\pm0.2\pi/a$, $N=50$, $J_1=6.52J_2$, $J_3=-3.22J_2$, $K=0.6J_2$, and $S=1.5$~\cite{PhysRevB.108.L180401,gn6c-1q19}.}
	\label{Fig2}	
\end{figure*}

In this Letter, we discover a unique anisotropy of the surface spin waves to serve as a signature of A-type altermagnets. Surface spin waves are collective excitations of spins localized at the surface of magnets~\cite{DEwave1,damon,camley1980long,wallis1967surface,PhysRev.185.752,PhysRev.185.762,PhysRevB.11.420,PhysRevB.60.15194,kostylev2013non,bkzn-qpqg}. It is a direct measurement of magnetic dynamics~\cite{demokritov2001brillouin,sebastian2015micro,tang2023spin} and is promising in spin-wave logic devices~\cite{chumak2015magnon,lenk2011building,jamali2013spin,PhysRevApplied.12.034015}. As shown in Fig.~\ref{Fig1}, the anisotropy of the surface spin wave in altermagnets manifests as a strong locking of the in-plane propagation direction, chirality, and top-bottom position. This anisotropic behavior is a result of the unique intrinsic $\mathcal{PT}$ symmetry breaking of altermagnets ($\mathcal{P}$ for spatial inversion, $\mathcal{T}$
for time reversal), thus is absent in conventional antiferromagnets and could serve as a compelling signature for detecting altermagnetism.
The broken $\mathcal{PT}$ symmetry arise directly from the nature of altermagnets, i.e., the spin-up
and spin-down sublattices cannot be related by translation or inversion.
We show that the anisotropic spin waves can be probed through the resonance  absorption spectrum ~\cite{ciubotaru2016all,gowtham2015traveling,maksymov2015broadband,li2017spin} or stray-field signal measured by, e.g., the nitrogen-vacancy center magnetometry~\cite{van2015nanometre,bertelli2020magnetic,melendez2025quantum}, thus establishes a unique pathway for characterizing altermagnetism as well as engineering functional devices based on altermagnets.





\textit{\textcolor{blue}{A-type altermagnet}.---}We consider an A-type altermagnet consisting of \( N \) layers of square lattices of out-of-plane spins \( \mathbf{S}_{i_l,j_l} \), where $l$ indexes the layers and \( (i_l, j_l) \) labels the intralayer lattice sites. For odd-numbered layers, the  spins feel the effective field \begin{eqnarray}\label{eq:Heff} 
\mathbf{H}_{\mathrm{eff}}^{(l)} &=& -\frac{2}{g \mu_0 \mu_B} \Big[ J_1 \big( \mathbf{S}_{i_l-1,j_l} + \mathbf{S}_{i_l+1,j_l} \big)+ J_2 \big( \mathbf{S}_{i_l,j_l-1}\notag \\ && +\mathbf{S}_{i_l,j_l+1} \big)+ J_3 \sum_{l'} \mathbf{S}_{i_{l'},j_{l'}} - K \big( \mathbf{S}_{i_l,j_l} \cdot \hat{\mathbf{z}} \big)\hat{\mathbf{z}} \Big], 
\end{eqnarray}
where \( J_1 > 0 \) and \( J_2 > 0 \) are the nearest-neighbor ferromagnetic exchange couplings along the $x$ and $y$-directions, respectively, \( J_3 < 0 \) is the interlayer antiferromagnetic exchange coupling, and \( K > 0 \) represents the anisotropy of the easy axis along the $z$ direction. 
For even-numbered layers, \( J_1 \) and \( J_2 \) are swapped. 
The combination $J_1>0,J_2> 0,J_3<0$
defines intralayer ferromagnetism and interlayer collinear antiferromagnetism. More importantly, when 
\begin{eqnarray}
J_1 \neq J_2,
\end{eqnarray}
the interlayer sublattices with opposite spins cannot be related through translation or inversion, so our model generically describes an A-type altermagnet. When \( J_1 = J_2 \), the system reduces to a conventional antiferromagnet. Later we will see that, $J_1\neq J_2$ not only defines the altermagnet, but also leads to the $\mathcal{PT}$ symmetry breaking, which gives rise to the anisotropy of the surface spin waves, thus the anisotropic surface spin waves can serve as a signature of the A-type altermagnets. In the case of out-of-plane N\'{e}el order, where conventional Damon-Eshbach surface spin waves are absent, the long-range dipolar interaction can be incorporated via a magnetodipolar contribution to \( K \) \cite{PhysRevB.43.6015}. For in-plane N\'{e}el order, the exchange-driven origin of the surface spin waves remains unchanged. The dipolar interaction introduces a magnetostatic correction to the dynamical matrix, which may quantitatively modify the dispersion, but it does not remove the exchange-induced surface-localized eigenmode. Therefore, 
our theory not only applies to A-type altermagnets 
with out-of-plane N\'{e}el order, such as $\text{CoNb}_4\text{Se}_8$~\cite{regmi2025altermagnetism,dale2024non,sakhya2025electronic}, CrSb~\cite{reimers2024direct,PhysRevLett.133.206401,zhou2025manipulation}, and twisted van der Waals antiferromagnets~\cite{cui2026altermagnetic} (e.g., twisted $\text{CrSP}_4$ or twisted $\text{CrI}_3$), but also works for A-type altermagnets with in-plane N\'{e}el order, such as $\text{MnTe}$~\cite{krempasky2024altermagnetic,PhysRevLett.132.036702,PhysRevB.109.115102} and FeS \cite{takagi2025spontaneous}.

\textit{\textcolor{blue}{Anisotropy of surface spin waves}.---}The surface spin waves can be solved by using the Landau–Lifshitz–Gilbert equation~\cite{ landau1935theory,gilbert2004}
\begin{equation}
\frac{d \mathbf{S}_{i_l,j_l}}{dt}
= -\gamma \mu_0 \mathbf{S}_{i_l,j_l} \times \mathbf{H}_{\mathrm{eff}}^{(l)}
+ \alpha_G \mathbf{S}_{i_l,j_l} \times \frac{d \mathbf{S}_{i_l,j_l}}{dt},
\label{eq:LLG}
\end{equation}
where \( \gamma = g \mu_B / \hbar \) is the gyromagnetic ratio, \( \mu_0 \) is the vacuum magnetic permeability, and \( \alpha_G \) is the Gilbert damping parameter. The effective magnetic field $\mathbf{H}_{\mathrm{eff}}^{(l)}$ has been given in Eq.~(\ref{eq:Heff}).
We proceed by linearizing the Landau–Lifshitz–Gilbert equation around the collinear magnetic ground state in the absence of damping (which is needed in the calculation in the resonance absorption spectrum later) and by assuming plane-wave excitations propagating within the $x-y$ plane. This leads to a characteristic eigenvalue problem for the spin-wave modes
$\hat{\eta} \, \mathcal{\hat{H}}_{\mathbf{q}} \, \mathcal{A} = \boldsymbol{\omega} \, \mathcal{A}$, where \( \mathcal{A} = (\mathcal{A}_1, \mathcal{A}_2, \ldots, \mathcal{A}_N)^T \) represents the vectors of the layer-resolved spin-wave amplitudes, and \( \hat{\eta} = \mathrm{diag}(1, -1, 1, -1, \ldots) \) is the metric that accounts for the interlayer antiferromagnetism. The dynamical matrix \( \mathcal{\hat{H}}_{\mathbf{q}} \) is tridiagonal with the matrix elements 
\begin{equation}
\mathcal{\hat{H}}_{\mathbf{q}}^{(i,j)} = 
\begin{cases} 
\omega_q + (s_l \cdot \omega_e) & \text{if } i = j, \\
\omega_e & \text{if } |i - j| = 1, \\
0 & \text{otherwise}.
\end{cases}
\label{eq:Hq}
\end{equation}
Here, \( \omega_q \) depends on the in-plane wave vector \( \mathbf{q} \) and \( \omega_e = -2 J_3 S / \hbar \) is the interlayer exchange contribution that couples adjacent layers, while the diagonal elements \( \omega_q + s_l \cdot \omega_e \) depend on the neighbor index \( s_l \), where \( s_l = 1 \) for the outermost layers and \( s_l = 2 \) for the inner layers. Such weakening of surface exchange coupling is essential for the emergence of the surface spin waves. The in-plane wave vector dependence enters through $\omega_q$, which differs between odd and even layers, i.e.,
$\hbar \omega_q^{(\mathrm{odd})} =  2KS  -  4J_1 S  (\cos q_x a - 1 ) -  4J_2 S (\cos q_y a - 1 )$, 
and
$\hbar\omega_q^{(\mathrm{even})} =  2KS -  4J_2 S (\cos q_x a - 1 ) -  4J_1 S  (\cos q_y a - 1 )$. Solving the spin-wave eigenvalue problem with the ansatz $m_l(\textbf{q},t)=\mathcal{A}_l e^{i(\textbf{q}r-\boldsymbol{\omega} t)}$ yields eigenfrequencies that separate into positive and negative branches, corresponding to opposite chiralities of precession. Adopting the convention that $\boldsymbol{\omega}>0$ ($\boldsymbol{\omega}<0$) labels right-handed (left-handed) chiralities, we can identify both right-handed and left-handed branches in the spectrum.

Figure~\ref{Fig2} presents the dispersions and amplitude profiles of the spin waves in the altermagnet. 
There are two surface spin waves (red and blue), gapped away from the bulk spin waves (gray) roughly by an energy of $\sqrt{J_3 K}$, where $J_3$ is the interlayer antiferromagnetic coupling and $K$ is the intralayer anisotropy energy. This gap is a property of the surface and bulk spin waves in antiferromagnets \cite{mills1968surface}. However, different from conventional antiferromagnets, the left-handed (blue) and right-handed (red) surface spin waves in the altermagnet show a splitting in frequency. 
This splitting is a result of the broken $\mathcal{PT}$-symmetry, thus is absent in conventional  antiferromagnets where degenerate surface modes can be simultaneously excited at any frequency ~\cite{sun2025exchange}. 
The amplitude profiles in the right panels of Fig.~\ref{Fig2} show that the left-handed and right-handed surface spin waves are localized on the top and bottom surfaces, respectively. For the N\'{e}el order in Fig.~\ref{Fig1}, the left-handed is on the top surface and right-handed is on the bottom surface. If the N\'{e}el order is reversed compared to that in Fig.~\ref{Fig1}, the top-bottom positions of the left- and right-handed modes are also swapped. This behavior of top-bottom positions is also due to the broken $\mathcal{PT}$ symmetry of altermagnet, thus is distinct from the Damon-Eshbach waves in ferromagnets or conventional antiferromagnets~\cite{DEwave1,damon,camley1980long,PhysRevB.28.1475,bkzn-qpqg}, where the top-bottom positions of the surface spin waves are swapped as the momentum is reversed due to nonreciprocity. 
The anisotropic nature of the surface spin waves is better seen in Fig.~\ref{Fig2}(c), where we show the equal-frequency contour of the two surface modes on the $q_x-q_y$ plane at the frequency marked by the dashed lines in Figs.~\ref{Fig2}(a) and \ref{Fig2}(b). 
Both surface spin waves show the elliptical equal-frequency contours, but the long axis of the contour of the left-handed (right-handed) mode is along the $q_x$ ($q_y$) direction. The chiral split equal-frequency contour 
is due to the lifting of the Kramer degeneracy, which is protected by either $\mathcal{PT}$ or $\mathcal{T}\tau$ symmetry, where $\tau$ is translation. Because the spin-opposite sublattices of altermagnets can not be related by $\mathcal{P}$ or $\tau$, the Kramer degeneracy is lifted, giving rise to the chiral split equal-frequency contour. 
The above chirality-dependent top-bottom positions and chiral split equal-frequency contours of the anisotropic surface spin waves are qualitatively different from those in conventional antiferromagnets, therefore provide signatures for identifying A-type altermagnets.

\begin{figure}
\includegraphics[width=0.48\textwidth]{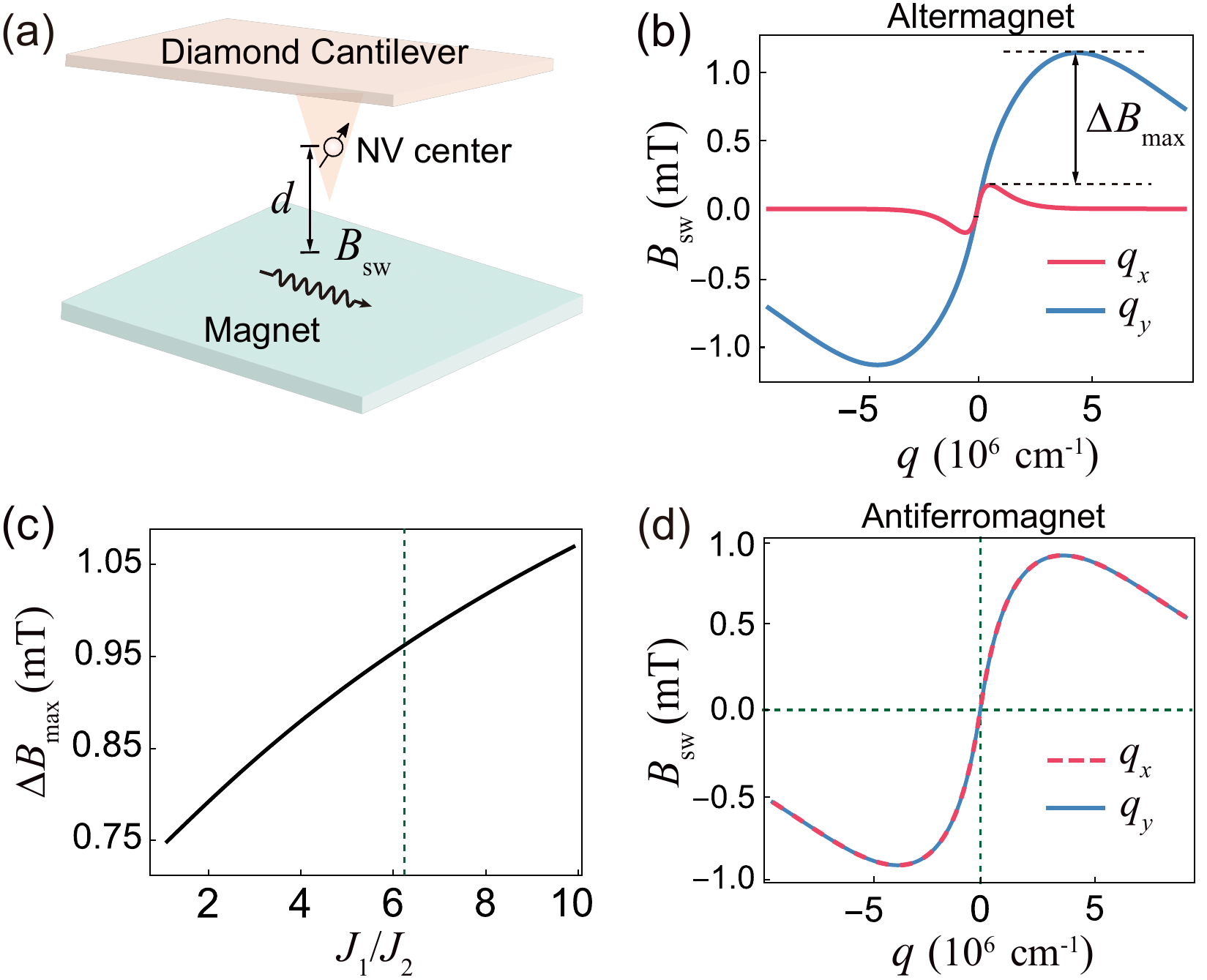}
	\caption{(a) Schematic of the experimental setup for measuring the stray field of the anisotropic surface spin wave using the nitrogen-vacancy (NV) center. (b) Stray field $B_{\text{sw}}$ of the surface spin waves in the altermagnet as a function of the wave vector $q$, $q=q_x$ ($q=q_y$) for the wave propagating along the $x$ ($y$) direction at $\mathbf{q}=(q_x,0)$ ($\mathbf{q}=(0,q_y)$). $\Delta B_\mathrm{max}$ is the difference between the maximum stray fields of surface spin waves propagating along the $x$ and $y$ directions. (c) $\Delta B_\mathrm{max}$ as a function of the ratio of two intralayer exchange couplings $J_1/J_2$. (d) The same as (b) but for antiferromagnet. Here, we set the total number of layers $N=30$, layer thickness $t_{0}=6.20$ \AA, distance between the nitrogen-vacancy center and surface $d=5t_{0}$, $J_1/J_2=6.22$ for the altermagnet and $J_1=J_2$ for the antiferromagnet.}
	\label{Fig3}	
\end{figure}

\textit{\textcolor{blue}{Stray field and absorption spectrum as experimental  signatures}.---} Now we show how the anisotropic surface spin waves in A-type altermagnets can be experimentally measured from the magnetic stray field or resonance absorption spectrum, by comparing them with conventional antiferromagnets. To have a realistic investigation, we focus on CoNb$_4$Se$_8$, a promising candidate for A-type altermagnets~\cite{regmi2025altermagnetism,dale2024non,sakhya2025electronic}, where the parameters are found by the first-principles calculations to be $J_1=J_2=5.80$ $\text{meV}$, $J_3=-25.60$ $\text{meV}$, $K=0.70$ $\text{meV}$,  $S=1$~\cite{regmi2025altermagnetism}, the lattice constant is $6.80\,\text{\AA}$, and the saturation magnetization $M_s$ is  $1.37 \mu_B$ per $\text{Co}$ atom. Although this earlier work does not distinguish the intralayer ferromagnetic exchange couplings $J_1$ and $J_2$, the chemical environment of non-magnetic atoms equivalently produces a $J_1  \neq J_2$ situation. In our treatment, we will fix $J_2=5.80$ $\text{meV}$ and consider different ratios of $J_1$ to $J_2$.

The magnetic stray field 
can by measured by, e.g., the nitrogen-vacancy center spectroscopy, as shown in 
Fig.~\ref{Fig3}(a). The stray field $B_{\text{sw}}$ can be simplified as~\cite{van2015nanometre}
\begin{equation}
B_{\text{sw}}(d, q) = \frac{1}{2} \mu_0 q t_0 M_s |\sum_{l=1}^{N}\mathcal{A}_{l}\exp[-|q||d-z_l|]|,
\end{equation}
where $d$ is the distance between the nitrogen-vacancy center in a diamond and sample surface,  $q$ is the wave vector of the spin waves along the $x$ or $y$ direction, $\mu_0$ is the vacuum magnetic permittivity, $t_0$ is the layer thickness, $M_s$ is the saturation magnetization, $\mathcal{A}_l$ is the amplitude at the $l$-th layer as those in Fig.~\ref{Fig2}, and $z_l$ is the height of the $l$-th layer, respectively. 

Figure~\ref{Fig3}(b) compares the stray fields for the right-handed surface spin waves along $q_x$ in Fig.~\ref{Fig2}(a) and the left-handed spin waves along $q_y$ in Fig.~\ref{Fig2}(b), because they are degenerate in frequency. 
A pronounced feature is that the stray field of the spin waves propagating along the $x$ direction is substantially weaker than those propagating along the $y$ direction. This behavior can be understood from Fig.~\ref{Fig2}, where the 
$x$-propagating surface spin waves are localized on the bottom surface, far away from the nitrogen-vacancy center that probes the stray field, while the $y$-propagating surface spin waves reside closer to the nitrogen-vacancy center. Consequently, for sufficiently thick sample, only the $y$-propagating surface spin waves can be probed. 
This anisotropy is due to the altermagnetic nature arising from 
$J_1\neq J_2$, so the difference in the stray field $\Delta B_\mathrm{sw}$ increases with the ratio of $J_1/J_2$, as shown in Fig.~\ref{Fig3}(c). By contrast, Fig.~\ref{Fig3}(d) show that, in  antiferromagnets the stray-field response is essentially identical for the two propagation directions, because the $x$- and $y$-propagating surface spin waves
are generated at both surfaces. 

\begin{figure}	\includegraphics[width=0.48\textwidth]{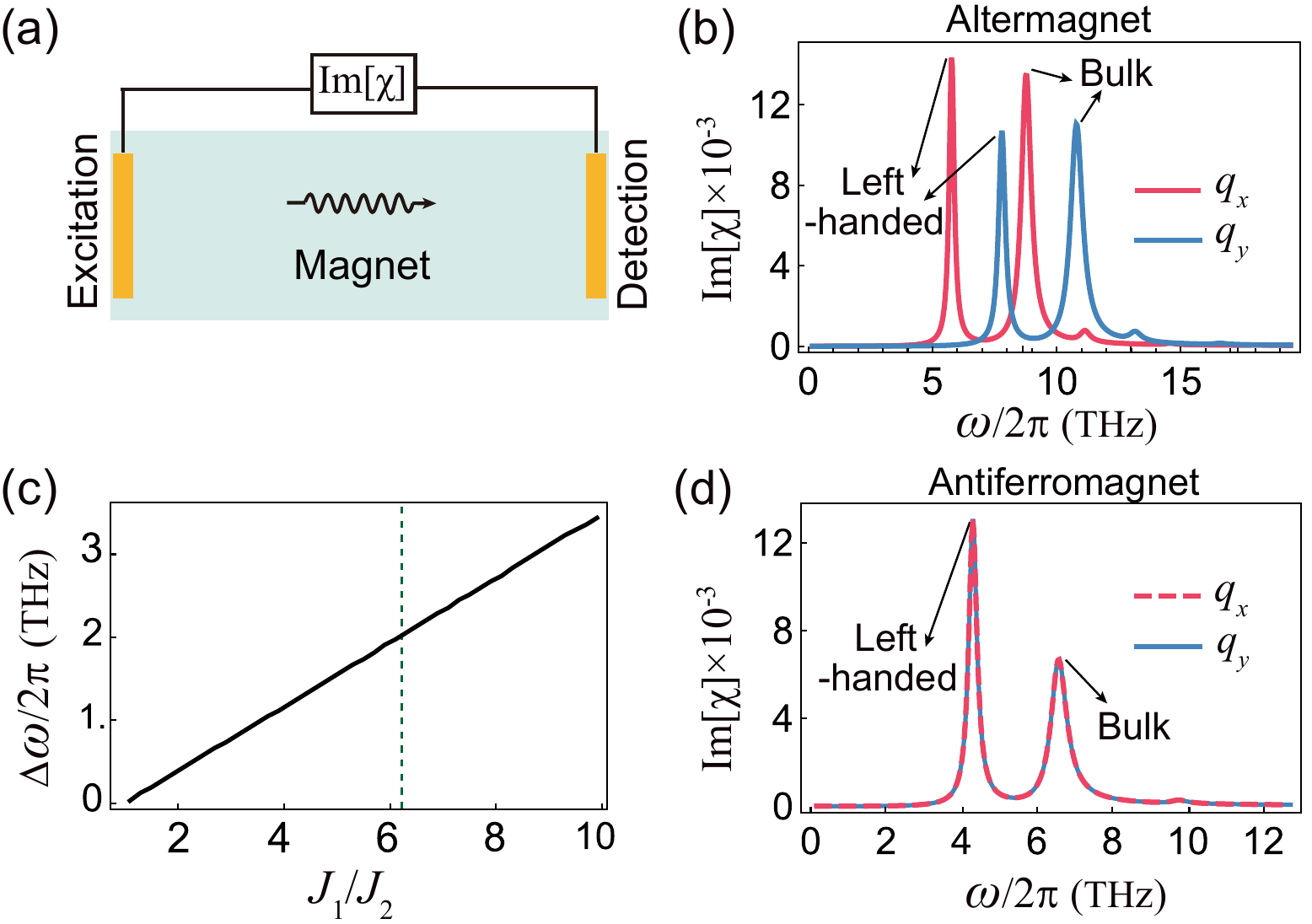}
	\caption{(a) Schematic of the  resonance absorption spectroscopy for measuring spin waves. Microwaves at a given wave vector $q$ can be excited by the excitation antenna and received by the detection antenna. A resonance absorption peak shows up at $\omega$ on the imaginary part of the total susceptibility $\mathrm{Im}[\chi]$, if $(\omega,q)$ matches the spin wave in the magnet.  (b) Left-handed circularly polarized absorption spectra for the altermagnet, given by  $\text{Im}[\chi]$ as a function of the frequency $\omega/2\pi$. (c) Frequency difference $\Delta\omega$ between two left-handed surface spin wave resonance absorption peaks as a function of the ratio of the intralayer exchange couplings $J_1/J_2$. (d) The same as (b) but for antiferromagnet. Here, we set the number of total layers $N=30$, Gilbert damping $\alpha_G=10^{-2}$, $h_0=1$ $\text{mT}$, $J_1/J_2=6.22$ for the altermagnet and $J_1=J_2$ for the antiferromagnet.}
	\label{Fig4}	
\end{figure}

The resonance absorption spectroscopy can also be used to measure the anisotropic surface spin waves. As shown in Fig.~\ref{Fig4}(a), microwaves at a given wave vector $q$ can be excited by the excitation antenna and received by the detection antenna and $q$ is determined by the distance between the antennas. A resonance absorption peak shows up at $\omega$, if $(\omega,q)$ matches a spin wave in the magnet. The power absorbed per unit area can be expressed as $\mathcal{P}_{\text{abs}}(\omega) = \omega t_0 h_0^2/\mu_0 \text{Im}\left[ \chi(\omega) \right]$, where \( t_0 \) is the layer thickness, \( h_0 \) is the microwave field amplitude, \( \mu_0 \) is the vacuum magnetic permeability, and the total susceptibility \( \chi(\omega) \) is given by~\cite{kittel1948theory}
\begin{equation}
\chi(\omega) = -\frac{\mu_0 M_s}{h_0} \sum_{n=1}^{N} \mathcal{A}_l(\omega),
\end{equation}
where \( \mathcal{A}_l(\omega) \) represents the spin-wave amplitude in the \( l \)-th layer and \( M_s \) is the saturation magnetization. 

As shown in Fig.~\ref{Fig4}(b), by exciting a left-handed circularly polarized wave with $q_x$ or $q_y$ = 5.5$\times 10^6$ cm$^{-1}$, 
the resonance absorption spectra $\mathrm{Im}[\chi]$ of the altermagnet shows noticeable split peaks, with 
the peaks associated with the $x$-propagating modes at lower frequencies than those propagating along the $y$ direction. This splitting can be traced back to the frequency splitting in Fig.~\ref{Fig2}, where the frequency of the left-handed surface spin waves is higher along $q_y$ than along $q_x$ for the same magnitude of wave vector. 
The bulk spin waves (the gray curves in Fig.~\ref{Fig2}) also show split resonance peaks at higher frequencies of Fig.~\ref{Fig4}(b), because their frequency spectrum in Fig.~\ref{Fig2} is also anisotropic. The bulk peaks decay rapidly with increasing frequency, because their uniform distribution leads to destructive interference. 
By contrast, Fig.~\ref{Fig4}(d) shows that there is no splitting in antiferromagnets. Therefore, the splitting in the resonance absorption spectrum can serve as a signature for the anisotropic surface spin waves in A-type altermagnets.


\textit{Conclusions.---} We have shown that A-type altermagnets can host a unique anisotropic surface spin waves, as a result of the altermagnetic nature that leads to broken $\mathcal{PT}$ symmetry. 
The anisotropic surface spin waves manifest as chirality-dependent top-bottom positions and chiral split equal-frequency contours, which are qualitatively different from those in conventional antiferromagnets, therefore providing signatures for identifying A-type altermagnets.
We further demonstrate that 
chirality-dependent top-bottom positions can be probed from the magnetic stray field using, e.g., the nitrogen-vacancy center spectroscopy, and the chiral split equal-frequency contours
can be measure by the resonance absorption spectroscopy. Our results establish surface spin waves as a sensitive probe of altermagnetism and will inspire magnonic devices based on altermagnets.

\textit{Acknowledgment.---}We thank Gerrit E. W. Bauer, Ping Tang, and Jilei Chen for valuable discussions. 

\bibliographystyle{apsrev4-1-etal-title_6authors}
\bibliography{reference}

\end{document}